\documentclass[fleqn,usenatbib]{mnras}

\usepackage{newtxtext,newtxmath}
\usepackage[T1]{fontenc}

\DeclareRobustCommand{\VAN}[3]{#2}
\let\VANthebibliography\thebibliography
\def\thebibliography{\DeclareRobustCommand{\VAN}[3]{##3}\VANthebibliography}

\usepackage{graphicx}	
\usepackage{amsmath}	
\usepackage{natbib}

\title[Bar stability in CDM and MOND]{A bar stability criterion distinguishing between modified gravity and dark matter in galaxies}

\author[T. Kashfi et al.]{
Tahere Kashfi,$^{1}$
Mahmood Roshan,$^{1}$\thanks{E-mail: mroshan@um.ac.ir}
Virginia Cuomo,$^{2}$
Benoit Famaey,$^{3}$
Asiyeh Habibi$^{1}$
and 
\newauthor
Srikanth T. Nagesh$^{3}$
\\
$^{1}$Ferdowsi University of Mashhad, P.O. Box 1436, Mashhad, Iran\\
$^{2}$Departamento de Astronomía, Universidad de La Serena,  Av. Raúl Bitrán 1305, La Serena, Chile\\
$^{3}$Universit\'e de Strasbourg, CNRS, Observatoire astronomique de Strasbourg, UMR 7550, F-67000 Strasbourg, France
}

\date{Accepted XXX. Received YYY; in original form ZZZ}

\pubyear{2015}

\begin{document}
\label{firstpage}
\pagerange{\pageref{firstpage}--\pageref{lastpage}}
\maketitle

\begin{abstract}
This paper presents a study on the distinguishability of dark matter and Modified Newtonian Dynamics (MOND) at galactic scales based on the stability criterion proposed by Efstathiou, Lake, and Negroponte (ELN criterion). First, we test the statistical validity of this stability criterion against the presence of bars within the SPARC and CALIFA databases, successfully identifying $\sim 70\%$ of barred galaxies. Then, we employ a series of N-body galaxy simulations to exhibit a direct observable difference between the dark matter and MOND theoretical frameworks, at least in gas-poor galaxies. We present N-body models that satisfy the stability requirement of the ELN criterion, and so are stable against bar formation in the presence of a dark matter halo, and that do actually exhibit bar instabilities in MOND. On the other hand, the question of how to inhibit bar formation in gas-poor galaxies in MOND is posed, and requires a detailed investigation of the external field effect.
\end{abstract}




\section{Introduction}

In the pioneering work by Efstathiou, Lake, and Negroponte \citep{Efs1982}, a simple but powerful criterion had been proposed for the global stability of exponential galactic discs -- in 2D gas-less N-body simulations including a rigid dark matter halo. The criterion (ELN criterion hereafter) introduces the stability parameter $\epsilon$ as
\begin{equation}\label{ELN}
	\epsilon=\frac{V_{\text{max}}}{\sqrt{G\, M_{d}/R_d }}
\end{equation}
where $V_{\text{max}}$, $M_d$, and $R_d$ refer to the maximum rotational velocity, the stellar disc mass, and the radial scale length of the exponential disc respectively. It suggests that galaxies with $\epsilon\leq 1.1$ are bar-unstable, while galaxies with $\epsilon>1.1$ are stable against bar formation. This criterion is based only on the global properties of the galaxy, which can be measured through observations, especially since gas-poor discs are often closer to exponential profiles than gas-rich ones \cite[see, e.g., the gas velocity profiles in][]{Lelli_2016}

More than two decades ago, the ELN criterion was applied by \citet{Mo1998} to develop a phenomenological model of disc galaxy formation within a hierarchical galaxy formation framework. Furthermore, semi-analytical models investigating the evolution of barred galaxies have consistently relied on the ELN criterion \citep{sem1,sem2}. However, there are obviously other factors that can influence bar formation and sustainability, such as gas cooling, feedback and galaxy mergers, or the spin of the dark matter halo. The latter has been shown to suppress bar formation in some studies \citep{halo1,halo2}, while others report the opposite behavior \citep{halo3}. The galactic bulge is another important factor, with high bulge-to-disc ratios suppressing bar formation in N-body simulations \citep{bulge1}, while observations have indeed shown that the bar fraction decreases with increasing bulge luminosity \citep{bulge2}. It is worth noting that this criterion needs modifications in certain N-body simulations, particularly when live halos are concerned. The shape of the halo velocity ellipsoid also affects the stability criterion \citep{Sellwood2016ApJ}. See \citet{Athanassoula2008} for other certain examples where the ELN criterion fails to distinguish bar stable from bar unstable discs. For a more recent study that presents a simulated galaxy with a Hernquist halo, where the ELN criterion is ineffective, we refer the reader to \cite{2024MNRAS.534..313F}.

However, the remarkable aspect is that despite not incorporating any of the aforementioned factors, the ELN criterion is still in relatively good agreement with $\Lambda$CDM cosmological simulations of galaxy formation at the statistical level. Indeed, the performance of the ELN criterion has recently been tested in the TNG50 simulation, and despite the inclusion of numerous complex physical mechanisms in this simulation, including feedback and galaxy mergers, this criterion successfully identifies approximately 75\% and 80\% of strongly barred and unbarred galaxies, respectively \citep{tng50}. The primary factor in the ELN criterion that appears to significantly influence the correct or incorrect classification of galaxies is the scale-length of the disc. Incorrectly classified barred galaxies (hence, classified as stable discs by the ELN criterion) tend to have much larger scale-lengths compared to typical barred galaxies, making them less compact. These galaxies are often embedded in dark matter halos with higher spin parameters, reflecting the positive correlation between scale length and spin parameter. Furthermore, at the epoch of bar formation, these simulated galaxies may have frequently undergone close encounters with massive satellites. This suggests that, for such galaxies, bar formation might be externally triggered rather than a natural outcome of the secular evolution of the disc. On the other hand, unbarred galaxies misclassified as bar-unstable by the ELN criterion typically have stellar discs similar to those of barred galaxies. However, their evolutionary histories differ: these discs assemble later than those of barred galaxies and are accompanied by significantly more prominent bulge components at earlier times. Furthermore, the vertical scale height of these discs is generally larger than in barred galaxies, indicating kinematically hotter discs. This suggests that additional environmental factors should be considered when determining disc stability \citep{tng50}. Nonetheless, the fact that the ELN criterion fails for approximately 20\% of the galaxies in TNG50 appears to be consistent with isolated simulations that have also reported failures of the ELN criterion. The surprising point is that it remains effective for the majority of galaxies in TNG50.

In this paper, we follow up on this by first evaluating the effectiveness of the ELN criterion for detecting barred galaxies in observations. Then, we demonstrate that galaxies with similar stability parameters $\epsilon$ and, as a result, similar physical observable properties $(R_d, M_d$,$V_{\text{max}})$, exhibit contrasting behavior in the dark matter paradigm compared to Modified Newtonian Dynamics (MOND) \citep{Milgrom_1983, Famaey_2012}. The paper is structured as follows: in Section \ref{sec:obs} we examine the validity of the ELN criterion in two observational samples; in Section \ref{sec:method}, we present isolated N-body simulations in both the dark matter and MOND frameworks. Additionally, a comprehensive comparison between the simulations in dark matter and MOND, with emphasis on the bar instability, is provided. Final comments and conclusions are discussed in Section \ref{sec:dis}.

\section{Observational implications}\label{sec:obs}

As outlined hereabove, the ELN criterion successfully identifies approximately 75\% and 80\% of strongly barred and unbarred galaxies, respectively, in the TNG50 cosmological simulation \citep{tng50}. Here, to assess the performance of this criterion on real observed galaxies, we explore two observational datasets. First, we use the Spitzer Photometry and Accurate Rotation Curves (SPARC) database \citep{Lelli_2016}, then we turn to a subsample of the Calar Alto Legacy Integral Field spectroscopy Area (CALIFA) survey \citep{Sanchez_2012}. 

SPARC is the largest galaxy sample which includes near-infrared ($3.6\mu m$) surface brightness and high-quality $\mathrm{HI/H\alpha}$ rotation curves. To investigate the effectiveness of the ELN criterion on the SPARC sample, we use the rotation curves provided by \cite{Li_2018}. We filter out galaxies with poor $\chi^2$ values exceeding 10, which might indicate various irregular behaviors. As mentioned in the previous section, galactic bulges can suppress the formation of bars. Therefore, to ensure the accuracy of our results, we excluded galaxies with a bulge from our samples. By focusing on bulgeless galaxies we refine our analysis to a selected sample of 132 galaxies. In this dataset, the identification of barred galaxies is based on the morphology types reported in the SPARC database. Note that this sample is biased against barred galaxies, and should not be used to infer the fraction of barred galaxies themselves (14/132 in the sample). Furthermore, we compute the stellar masses of the galaxies using their luminosities and mass-to-light ratios provided by \cite{Li_2018}. The physical properties of these galaxies are listed in Table~\ref{tab:table1}.

Within this sample, we find that the ELN criterion is smaller than 1.1 ($\epsilon\leq 1.1$) for $71\%\pm12\%$ of the identified barred galaxies. On the other hand, 81\%$\pm3.6$\% of the unbarred galaxies included in the selected sample are characterized by $\epsilon > 1.1$. The uncertainties of the percentages are calculated using the formula for the standard error of a proportion, $\sqrt{p(1-p)/n}$, where $p$ is the proportion and $n$ is the sample size. These results show that, the ELN criterion works very well observationally both to identify barred and unbarred galaxies. \footnote{Additionally, when considering the baryonic (stars+gas) mass in equation \eqref{ELN}, since the value of $\epsilon$ automatically decreases, we find that 92.8\% of barred galaxies in the SPARC dataset have an ELN criterion smaller than 1.1, but that only 37\% of unbarred galaxies then have an ELN criterion larger than 1.1.}

\begin{table*}
	\centering
	\caption{The physical and observational properties of barred galaxies in the SPARC dataset. The columns contain (1): the main ID of each galaxy (2): stellar mass in units of $10^{10}\ M_{\odot}$ (3): the radial scale length (kpc) (4): the maximum rotational velocity (km/s)  (5): the ELN criterion.}\label{tab:table1}
		\begin{tabular}{ccccc}
			\hline
			Galaxy name & $M_d$ & $R_d$ & $V_{\text{max}}$ & $\epsilon$ \\
			
			& ($10^{10}M_{\odot}$) & (kpc) & (km/s) & \\ 
			\hline
			NGC1090 & $5.33\pm0.56$ & $2.33\pm0.75$ & $176.76\pm7.38$ & $0.56\pm0.14$ \\			
			NGC4088 & $4.29\pm0.77$ & $1.91\pm0.45$ & $182.74\pm6.12$ & $0.58\pm0.14$ \\
			NGC4100 & $4.51\pm0.62$  & $1.78\pm0.39$ & $195.31\pm6.11$ & $0.59\pm0.12$ \\
			NGC3992 & $17.25\pm2.33$ & $4.46\pm0.80$ & $274.95\pm5.76$ & $0.67\pm0.12$ \\
			NGC5055 & $8.56\pm0.17$ & $3.17\pm0.19$ & $243.80\pm0.50$ & $0.71\pm0.03$ \\			
			NGC4051 & $4.29\pm0.88$ & $3.95\pm1.04$ & $165.87\pm5.36$ & $0.77\pm0.20$ \\
			NGC3953 & $8.33\pm1.44$ & $4.35\pm1.09$ & $224.21\pm6.10$ & $0.78\pm0.19$ \\
			NGC0289 & $6.63\pm0.69$ & $4.78\pm2.13$ & $205.01\pm33.81$  & $0.84\pm0.37$ \\
			NGC3521 & $3.90\pm0.43$ & $2.66\pm1.05$ & $219.69\pm14.58$ & $0.87\pm0.28$ \\
			NGC4389 & $0.64\pm0.16$ & $1.78\pm0.59$ & $124.02\pm8.57$ & $1.00\pm0.35$ \\			 
			NGC3972 & $0.72\pm0.12$ & $2.33\pm0.53$ & $134.00\pm4.90$ & $1.16\pm0.27$ \\
			NGC3949 & $1.67\pm0.28$ & $3.45\pm0.86$ & $169.41\pm25.56$ & $1.17\pm0.42$ \\
			NGC3769 & $0.76\pm0.14$ & $3.04\pm0.67$ & $126.16\pm4.40$ & $1.21\pm0.28$ \\					
			ESO079-G014 & $2.59\pm0.49$ & $5.59\pm1.93$ & $177.76\pm2.03$ & $1.26\pm0.35$ \\
			\hline
		\end{tabular}
\end{table*}

These percentages based on the SPARC database are nevertheless relatively crude, due to the fact that bars are only identified visually in this sample and that the SPARC sample is biased against barred galaxies by favoring highly symmetric two-dimensional velocity fields for high-quality rotation curves. So, as an additional observational appraisal of the ELN criterion, we apply it to the subsample of the CALIFA survey analyzed by \cite{Aguerri_2015, Kalinova_2017, Cuomo_2020}. We select disc galaxies with flat rotation curves and derive the maximum rotational velocities using an asymmetric drift correction, as done by \cite{Cuomo_2020}, or calculating the mean value at radii larger than the effective radius of the circular velocity curves presented by \cite{Kalinova_2017}, after checking that the two approaches produce maximum velocity values that are consistent for the galaxies analyzed in both of the mentioned studies. Moreover, we select galaxies dominated by the disc, with a disc-to-total luminosity ratio $D/T$ larger than 0.9, according to the values obtained through the photometric decomposition performed using SDSS $r$-band images presented by \cite{MendezAbreu_2017}. The same photometric decomposition provides us the disc scale length as well. We select a final sample of 20 galaxies, which includes 15 barred ones. Table~\ref{tab:table2} indicate the physical properties of this sample. 

\begin{table*}
	\centering
	\caption{The physical and observational properties of barred galaxies in the CALIFA survey. The columns contain (1): the main ID of each galaxy (2): stellar mass in units of $10^{10}\ M_{\odot}$ (3): the radial scale length (kpc) (4): the maximum rotational velocity (km/s)  (5): the ELN criterion.}\label{tab:table2}
		\begin{tabular}{ccccc}
			\hline
			Galaxy name & $M_d$ & $R_d$ & $V_{\text{max}}$ & $\epsilon$ \\
			
			& ($10^{10}M_{\odot}$) & (kpc) & (km/s) & \\ 
			\hline
			NGC0776 & $6.61$ & $5.47\pm0.23$ & $138.62\pm0.00$ & $0.61\pm0.20$ \\
			NGC3994 & $2.57$ & $1.41\pm0.06$ & $2246.40\pm5.50$ & $0.81\pm0.04$ \\
			NGC6941 & $9.77$ & $7.38\pm0.31$ & $196.60\pm0.00$ & $0.82\pm0.01$ \\
			NGC7321 & $10.71$ & $5.39\pm0.23$ & $254.90\pm0.00$ & $0.87\pm0.01$ \\												
			IC5309 & $1.32$ & $3.42\pm0.29$ & $113.60\pm24.50$ & $0.88\pm0.23$ \\
			UGC03253 & $3.31$ & $3.60\pm0.15$ & $184.40\pm0.00$ & $0.93\pm0.01$ \\
			NGC3811 & $2.57$ & $3.34\pm0.10$ & $180.65\pm0.00$ & $0.99\pm0.16$ \\
			NGC6060 & $8.13$ & $6.77\pm0.19$ & $225.10\pm0.00$ & $0.99\pm0.26$ \\
			NGC192 & $5.01$ & $3.92\pm0.11$ & $248.30\pm6.60$ & $1.06\pm0.04$ \\												
			NGC551 & $3.89$ & $4.93\pm0.21$ & $202.00\pm42.70$  & $1.10\pm0.25$ \\
			IC1528 & $1.23$ & $3.86\pm0.25$ & $141.80\pm13.70$ & $1.21\pm0.16$ \\			
			NGC5205 & $0.63$ & $1.89\pm0.08$ & $170.60\pm0.00$ & $1.42\pm0.02$ \\			
			UGC3944 & $0.83$ & $3.33\pm0.15$ & $147.80\pm30.10$ & $1.42\pm0.32$ \\			
			MCG0202030 & $1.95$ & $4.12\pm0.18$ & $210.00\pm55.50$ & $1.47\pm0.42$ \\
			UGC8231 & $0.13$ & $2.40\pm0.11$ & $135.90\pm27.20$ & $2.83\pm0.63$ \\ 
			\hline
		\end{tabular}
\end{table*}

To examine the validity of the ELN criterion in this subsample, we adopt the total stellar mass values derived by \cite{GonzalezDelgado_2015} and find that $67\%\pm12\%$ of barred galaxies have $\epsilon\leq 1.1$, very similar to the fraction of barred galaxies identified by the ELN criterion in SPARC. While our investigation in the CALIFA survey does not encompass a large sample of galaxies, our findings demonstrate the capability of the ELN criterion to be applied to real galaxies. By establishing a similarity between the results obtained from the TNG50 cosmological simulation and these two observational databases, we can conclude that the ELN criterion serves as a simple and viable criterion for statistically detecting barred galaxies, achieving an approximate success rate of 70\%.
\footnote{As an additional analysis, we also examined data from the Mapping Nearby Galaxies at Apache Point Observatory (MaNGA) survey \citep{MaNGA1}. This survey includes a sample of approximately 180 barred galaxies, 46 of which are suitable for applying the Tremaine-Weinberg method to measure the pattern speeds of barred galaxies \citep{MaNGA2, Geron2023}. Our analysis showed that the ELN criterion is consistently smaller than 1.1 for {\it all} these galaxies, further supporting the efficiency of the ELN criterion.}

\begin{figure}
	\centering
	\includegraphics[width=1.0\linewidth]{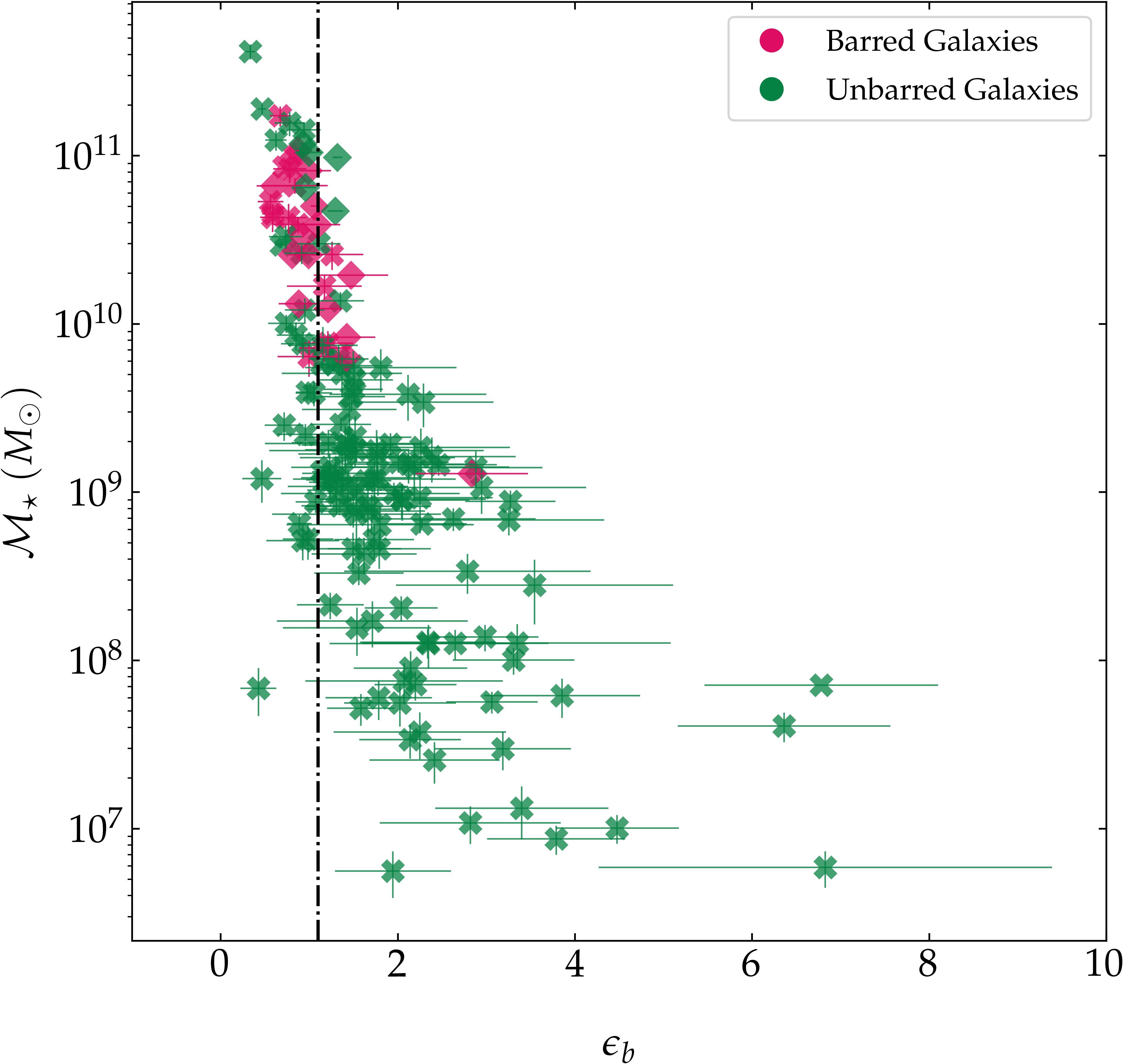}
	\caption{Stellar mass versus the stability criterion for all galaxies in SPARC (filled X) and CALIFA (diamond). Barred and unbarred galaxies are shown by red and green points, respectively. The vertical dashed line indicates the stability threshold.}\label{fig:m-epsilon}
\end{figure}

Figure~\ref{fig:m-epsilon} displays the stellar mass versus the stability criterion for all galaxies in the two observational samples. In this figure Barred and unbarred galaxies are represented by red and green points, respectively. The vertical dashed line denotes the ELN stability cutoff. As evident from the figure, the criterion effectively identifies where barred galaxies lie in both datasets. Notably, it performs particularly well in identifying unbarred galaxies in the SPARC dataset, where the fraction of correctly classified unbarred galaxies is higher. Note that there is a significant fraction of unbarred galaxies with a low ELN parameter within SPARC, but the relative numbers are not meaningful since the dataset is biased against selecting barred galaxies.

Drawing implicit dependence on the gravitational force, the ELN criterion actually also holds the potential to serve as a means for testing different gravitational theories. To make a first step in the direction of exploring this possibility, the following section presents numerical simulations of gas-less isolated galaxies within two different gravitational frameworks. Through these simulations, we aim to examine how the ELN criterion behaves when subjected to variations in the underlying gravitational models. 

\section{Isolated N-body Simulations in dark matter and MOND} \label{sec:method}

In this Section, we present a series of numerical simulations aimed at studying the stability of stellar discs against the bar instability within the context of the dark matter and MOND paradigms. Our objective is to decipher the meaning of the ELN stability criterion within these two paradigms.

Our dark matter simulations consist of two components: a spherical dark matter halo and a stellar disc. For these simulations, we employ the \textsc{RAMSES} code \citep{RAMSES_2002}, which uses the AMR (Adaptive-Mesh-Refinement) technique, increasing grid resolution in high-density regions without significantly increasing calculation time. For our single component disc simulations in MOND, we use the Phantom of \textsc{RAMSES} (\textsc{POR}) patch of the \textsc{RAMSES} code \citep{Fabian_2015,Nagesh_2021}. For a concise documentation of MOND simulations thus far, and the POR package accompanied with a manual, we refer the reader to \cite{Nagesh_2021}. For recent galactic simulations with POR see \cite{Banik_2020} and \cite{Roshan_2021}.

\subsection{Initial conditions}\label{IC}

All our models incorporate an exponential disc for the stellar component, which is described by the following density profile
\begin{equation}\label{disk_den}
	\rho(R,z)=\frac{M_d}{4\pi z_d R_d^2} e^{-\frac{R}{R_d}} \text{sech}^2\left(\frac{z}{2z_d}\right)
\end{equation}
where $M_d$ is the disc mass and $R_d$ and $z_d$ are the scale length and scale height of the disc, respectively. The initial conditions for MOND simulations are generated with disc Initial Conditions Environment (DICE) \citep{Banik_2020}. Because the original version of DICE is not compatible with MOND, we use the modified version of it which is publicly available \footnote{https://bitbucket.org/SrikanthTN/bonnpor/src/master/}. 

Since the ELN criterion depends on the mass of the disc and its radial scale length, we implement three stellar discs with different characteristics. Specifically, we construct three models with stellar masses of $10^{11}M_{\odot}$, $10^{10}M_{\odot}$, and $10^{9}M_{\odot}$ which we refer to as MOND1, MOND2, and MOND3 respectively. Then we fix the scale length of these models using the following $R_d-M_{*}$ relation~\citep{Pizagno_2005},
\begin{equation}\label{R_Mass}
	R_d=3.87(\pm 0.11)\big(\frac{M_*}{2.21\times10^{10}M_{\odot}}\big)^{0.24\pm 0.03}\,\,\,\, \text{kpc},
\end{equation}
where in our case the stellar mass $M_*$ is equal to $M_d$. We choose a typical value for the scale height $z_d$ so that the disc flattening $R_d/2z_d$ in our models falls within the average interval of $7.3\pm 2.2$ \citep{Kregel_2002}. 

On the other hand, for our Newtonian simulations, we adopt the Navarro, Frenk, \& White (NFW) halo, which is represented by the following density profile \citep{NFW}:
\begin{eqnarray}\label{halo_den}
	\rho_{\text{NFW}}(r)=\frac{M_h}{4\pi r_h^3}\Bigg[\frac{r}{r_h}\Big(1+\frac{r}{r_h}\Big)^2\Bigg]^{-1},
\end{eqnarray}
with $M_h$ and $r_h$ denoting the mass and scale radius parameters. Although we have also conducted simulations using the Plummer halo, the primary advantage of the NFW halo is its consistency with the dark matter mass distributions observed in cosmological simulations. 

The total rotation curve of the Newtonian disc-halo models is given by the addition in quadrature of the disc and halo rotation curves
\begin{equation}\label{velociy}
	V^2(x)=\frac{M_d G}{2 R_d}\Big[x^2\mathcal{V}(\frac{x}{2})+6\beta_2\frac{(x+\beta_1)\ln(1+\frac{x}{\beta_1})-x}{x(x+\beta_1)}\Big]
\end{equation}
where $x=\frac{r}{R_d}$, $\beta_1=\frac{r_h}{R_d}$, $\beta_2=\frac{M_h}{M_d}$, and the function $\mathcal{V}$ is related to the modified Bessel functions as $\mathcal{V}=I_0 K_0-I_1 K_1$. By examining the rotation curve, it becomes evident that the ELN criterion solely depends on $\beta_1$ and $\beta_2$, i.e., $\epsilon=\epsilon(\beta_1,\beta_2)$. To make the initial conditions for dark matter simulations we use the \textsc{GALAXY} code \citep{Sellwood_2014_galaxy}. In this code, we just need to set the value of $\beta_1$, $\beta_2$ and $\beta_3=\frac{z_d}{R_d}$. The latter does not contribute to the ELN criterion. By solely setting the ratio of halo to disc properties, our models effectively represent an infinite number of galaxies. In order to resolve this degeneracy, we enforce adherence to the $c-M_{200}$ relation \citep{cm} in our models. Here, $c=r_{200}/r_h$ represents the halo concentration parameter, $r_{200}$ denotes the radius at which the halo density is 200 times the critical density of the universe $\rho_{\text{crit}}$, and $M_{200}$ is defined as $M_{200}=(4\pi/3) 200\rho_{\text{crit}} r_{200}^3$. We construct our dark matter models by adopting the same disc parameters as those used in the MOND models. Then for a given halo mass parameter $M_h$ (or equivalently $\beta_2$), we specify the halo scale radius $r_h$ (or equivalently $\beta_1$) using the $c-M_{200}$ relation \citep{niro}:
\begin{eqnarray}\label{rh_M200}
	r_h\simeq 28.8 \left(\frac{M_{200}}{10^{12} h^{-1}M_{\odot}}\right)^{0.43} \, \text{kpc}
\end{eqnarray}
It should be noted that the \textsc{GALAXY} code utilizes adiabatic compression to bring the halo into equilibrium with the embedded exponential disc \citep{compress}.

After the initial conditions are set by \textsc{GALAXY}, we run our dark matter simulations with \textsc{RAMSES}\footnote{Of course, we also ran the simulations using the \textsc{GALAXY} code, and the results were qualitatively identical. The only reason we conducted the dark matter simulations with \textsc{RAMSES} is to ensure consistency by using the same code for both MOND and dark matter simulations.}. In the end, we devise three models in Newtonian gravity, which we name NFW1, NFW2, and NFW3. To prevent local instability in our models, we have set the Toomre's stability parameter (see \cite{Toomre_1964}) to $Q=1.5$ at $t=0$. We tracked the evolution of the discs with $N=1\times 10^6$ equal mass particles over a period of 7.5 Gyr. Table~\ref{tab:table3} briefly highlights some key parameters, including the initial value of the ELN criterion, of these simulations.

\begin{table}
	\centering
		\caption{Simulations parameters. The columns contain (1): acronym used for simulation models (2): initial disc mass in units of $10^{10}\ M_{\odot}$ (3): the radial scale length (kpc) (4): the ratio of halo's radial scale length over $R_d$ (5): halo's mass scaled by $M_d$ (6): halo's truncation radius scaled by the radial scale length of the halo (7): halo's concentration parameter (8): the initial ELN criterion.}\label{tab:table3}
		\begin{tabular}{ccccccccc}
			\hline
			Simulation & $M_d$ & $R_d$ & $\beta_1$ & $\beta_2$ & $\frac{r_{\text{trun}}}{r_h}$ & $c$ & $\epsilon$ \\
			
			Model & ($10^{10}M_{\odot}$) & (kpc) &  &  &  &  &  \\ 
			\hline
			NFW1 & 10.0 & 5.56  & 1.26 & {2.76} & 15 & {11.06} & 0.93 \\
			NFW2 & 1.0 & 3.20 & 1.22 & {6.90} & 15 & {12.60} & 1.20  \\
			NFW3 & 0.1 & 1.84 & 1.12 & {15.67} & 15 & {14.50} & 1.63 \\

			\hline
			MOND1 & 10.0 & 5.56 & -- & -- & -- & -- & 0.85 \\
			MOND2 & 1.0 & 3.20 & -- & -- & -- & -- & 1.10 \\
			MOND3 & 0.1 & 1.84 & -- & -- & -- & -- & 1.40 \\
			\hline
		\end{tabular}
\end{table}

\begin{figure}
	\centering
	\includegraphics[width=1.0\linewidth]{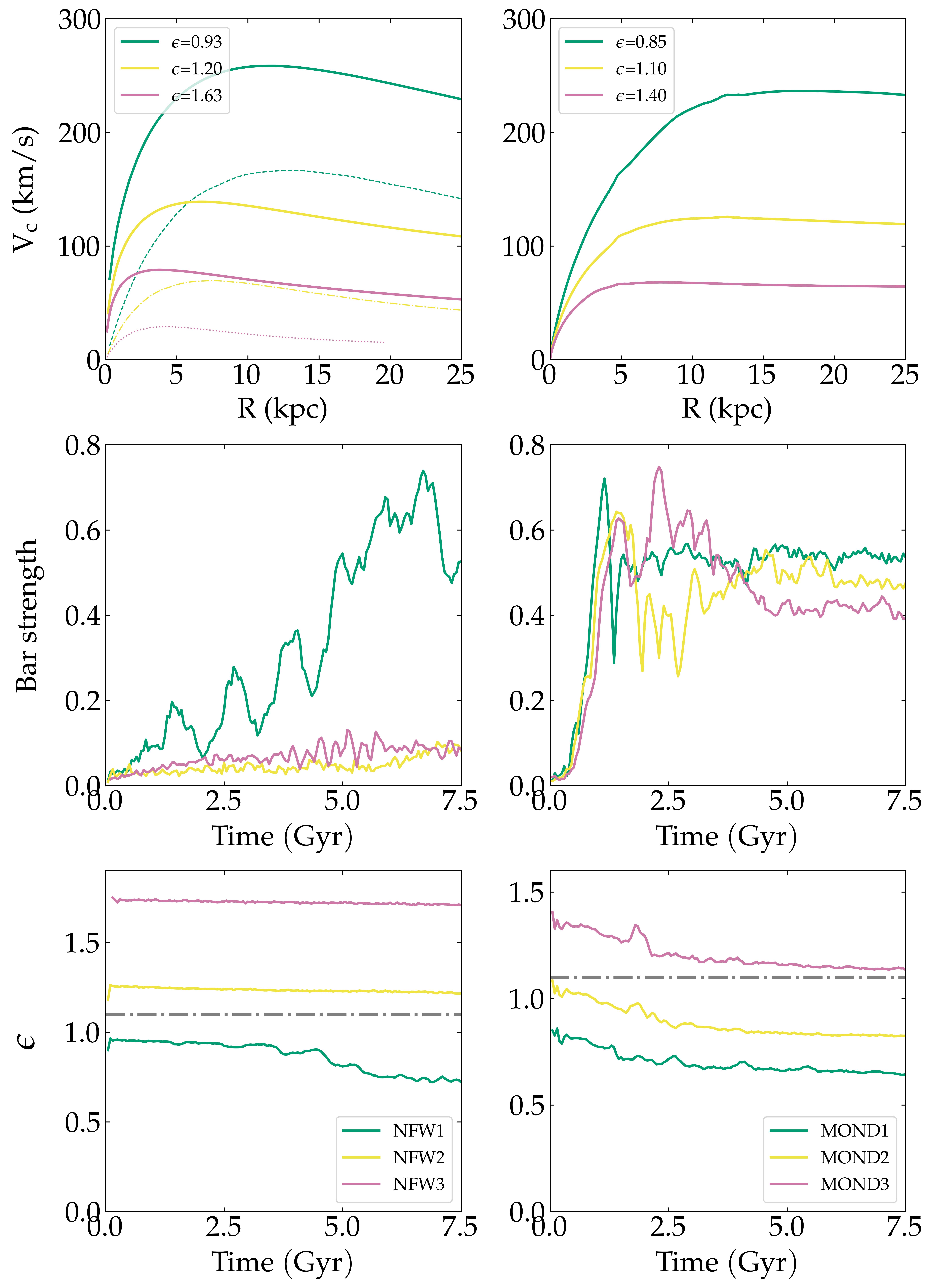}
	\caption{\textit{Top row}: initial rotational velocities for the NFW (left) and MOND (right) models. Discs with different ELN criteria are illustrated with different colours. The Newtonian contribution of the disc in NFW models are shown by dashed, dashdot and dotted lines. \textit{Middle row}: the time evolution of the corresponding bar amplitudes in the models. \textit{Bottom row}: the time evolution of ELN criterion. The dash-dot lines indicate the stability cut-off. In both middle and bottom rows, the left panel corresponds to the dark matter model, while the right panel corresponds to the MOND model.}\label{fig:rc-ampl}
\end{figure}

The top panels in Fig.~\ref{fig:rc-ampl} depict the rotation curves of our models. Specifically, the top left and top right panels correspond to simulations based on MOND and dark matter, respectively. As expected, the rotation curves from both approaches exhibit a good degree of similarity. 

\subsection{Simulations outputs}\label{sec:results}

Having described the initial state of the simulations, we now explore the occurrence of the bar instability in accordance with our main objective. The bar strength is commonly defined in terms of the Fourier decomposition of the mass distribution as $A_2^{\text{max}}=\max[A_2(R)]$.
Here $A_2(R)$ is the Fourier amplitude of mode $m=2$ at radius $R$ \citep{Algorry_2017, Guo_2019}. For arbitrary mode $m$ the amplitude is written in terms of Fourier coefficients
\begin{equation}\label{a2}
	A_m(R)\equiv \sqrt{a_m(R)^2+b_m(R)^2},
\end{equation}
where
\begin{eqnarray}\label{coefficients}
	&a_m(R)=\frac{1}{N} \displaystyle\sum_{k=1}^{N}\cos m\phi_k,\  m=1,2,...\\
	&b_m(R)=\frac{1}{N} \displaystyle\sum_{k=1}^{N}\sin m\phi_k,\  m=1,2,...
\end{eqnarray}
To compute these coefficients, we divide the disc into annuli of equal width and consider only disc particles with $|z|<1$ kpc to avoid the discreteness noise caused by the particles which have higher vertical distances. In these equations, $N$ represents the number of particles in each annulus. Conveniently, $A_2^{\text{max}}\geq 0.4$ corresponds to strong bars, $0.2\leq A_2^{\text{max}}< 0.4$ indicates weak bars, and the unbarred discs are characterized by $A_2^{\text{max}}<0.2$~\citep{Algorry_2017}.

\begin{figure} 
	\centering
	\includegraphics[width=1.0\linewidth]{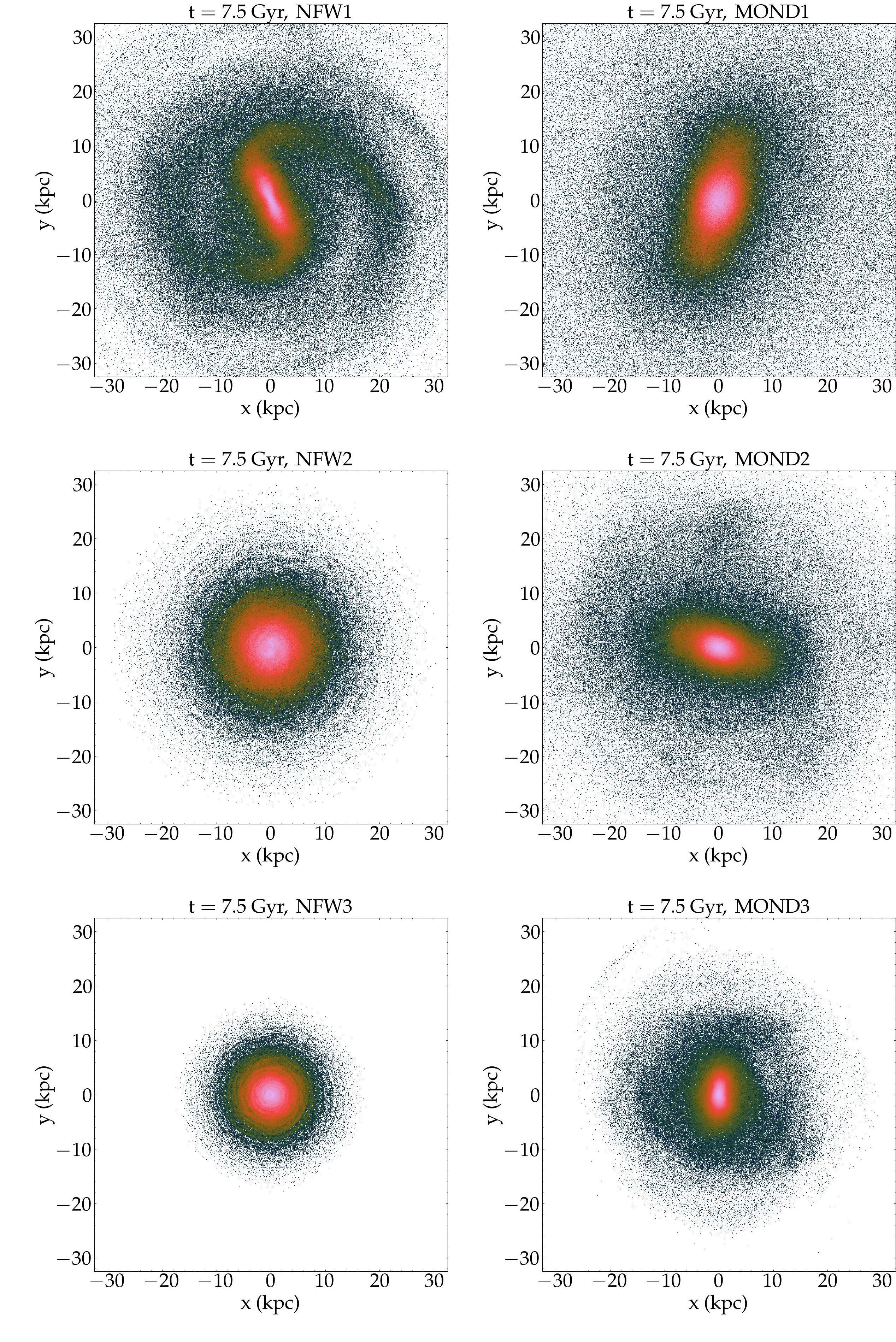}
	\caption{ The face on view of the discs at the end of the simulation. The first and second columns correspond to the dark matter and MOND models, respectively.}\label{face}
\end{figure}

We begin by examining the time evolution of bar amplitudes in our dark matter models. From the middle panel of Fig.~\ref{fig:rc-ampl}, it is evident that the bar amplitude gradually increases only in the NFW1 model with $\epsilon<1.1$, ultimately giving rise to a strong bar over a period of 5 Gyr. Consistent with the ELN criterion, the other two models with $\epsilon>1.1$ remain entirely stable against bar formation. As previously noted, this behavior has been recognized for more than four decades. Fig.~\ref{face} displays the face-on view of the disc at the end of the simulation. The dark matter models are presented in the right column. In the NFW1 model, a prominent bar is visible along with spiral arms.

In MOND models, a distinct behavior emerges. As illustrated in the middle left panel of Fig.~\ref{fig:rc-ampl}, although increasing the ELN criterion results in a (barely noticeable) reduction in the initial bar growth rate, it does not entirely eliminate bar instability. Specifically, while the MOND1 model with the smallest $\epsilon$ exhibits the strongest bar at the end of the simulation, both MOND2 and MOND3, with initial $\epsilon \geq 1.1$, harbor strong bars as well. In the MOND3 case, $\epsilon$ even remains above 1.1 for the whole duration of the simulation. The strong bars are visible in the face on view of the MOND discs shown in the left column of Fig.~\ref{face}.

The time evolution of the ELN criterion, $\epsilon$, is displayed in the bottom panels of Fig.~\ref{fig:rc-ampl}. In stable models NFW2 and NFW3, this parameter remains constant. However, in the other unstable models, $\epsilon$ decreases during the formation of the bar and the buckling instability. This can be attributed mainly to the decrease in $R_d$ during the bar formation.
We highlight again that our dark matter simulations have been duplicated using the \textsc{GALAXY} code, and the resulting outcomes are identical.

It is worth noting that we created additional models with varying properties, particularly those with a smaller range of concentration parameters. While these models may not fully conform to the $c-M_{200}$ relation, they still demonstrate that MOND-simulated galaxies violate the ELN criterion, even with different initial conditions.

\section{Discussion and conclusion}\label{sec:dis}

In conclusion, from a statistical perspective, the ELN criterion of Eq.~\eqref{ELN} proves to be an effective criterion for {\it statistically} detecting barred galaxies in both standard cosmological simulations and in observations, as we demonstrated here (Section~2) with small but up-to-date datasets. However, this criterion has a non-negligible `false positive' and `false negative' rates, and we point out that it is imperative to study cases where this criterion does not apply, as such cases could in principle provide valuable insights into the stability of galaxies and bar formation under different gravitational scenarios.

Our (gas-less) numerical simulations (Section~3) indeed show that bars form much more readily in MOND than in the dark matter case \citep{Tiret_2007}, particularly within gas-poor galaxies. On the one hand, as it is expected, galaxies with large $\epsilon$ values demonstrate stability within dark matter simulations, but on the other hand, in the MOND model, all stellar discs are self-gravitating and gas-poor disc galaxies seem to all have the propensity to host bars. It is however important to understand the limitations of our simulations, as they do not account for the presence of gas. Without including gas in these simulations, any definitive conclusions would be premature. For a comprehensive study of including the gas component in MOND simulations, we refer the reader to \citet{Nagesh_2023}, and more systematic investigations should be carried out. Bar weakening due to gas inflows are indeed known to be important both in the standard and in the MOND context \citep{Tiret_2008}. The existence of gas-poor unbarred galaxies may however present a challenge to the MOND model, although it should be noted that even a weak external field effect \citep[][]{Banik_2020} might also change the results in the MOND context.

While our dark matter simulations are, on the other hand, broadly consistent with observations, as shown in Section~2, there still is a $\sim 30$\% fraction of barred galaxies that harbour an ELN criterion larger than 1.1. Conversely, our findings highlight the potential of MOND in explaining the existence of such barred galaxies with a large ELN criterion.  Therefore, it is interesting to look for barred gas-poor galaxies with a large ELN criterion in observational datasets. Some notable examples come from the analysis of the CALIFA galaxies presented in Sect.~\ref{sec:obs}. The first interesting case is NGC5205, a barred galaxy located in the Ursa Major constellation, observed as part of the CALIFA survey and analysed by \citet{Aguerri_2015}. Indeed, NGC5205 is an intriguing case as it manifests a gas fraction smaller than 0.05 \citep{GarmaOehmichen_2020}, and a large ELN criterion value of $\epsilon = 1.48 \pm 0.18$. This observation piques our interest and suggests that MOND may provide a plausible explanation for the existence of bars in such galactic systems. Additionally, we have identified two other candidates in the CALIFA survey -- UGC8231 and UGC3944 -- included in the analysis presented by \citet{Cuomo_2020}, both of which exhibit large ELN criteria of $\epsilon = 2.83\pm0.63$ and $\epsilon = 1.42\pm0.32$, respectively. These cases where MOND might potentially provide an easier way to explain observations are also in line with the observations of barred submaximal discs, which had been previously shown to be more numerous in observations than in TNG50 \citep{Kashfi_2023}.

In summary, the coexistence of barred and unbarred systems, both displaying large values of $\epsilon$, prompts a need for a deeper understanding of the underlying dynamics in both gravitational frameworks. Moreover, our analysis has highlighted the presence of certain gas-poor barred galaxies in observational samples, exhibiting a large ELN criterion, which might present a challenge to our understanding of stability against bar formation in the standard context, whilst the observed gas-poor unbarred galaxies might present a challenge to MOND. However, further analysis, considering in particular gas dynamics and the external field effect, is required for a more definitive statistical conclusion, which could ultimately lead to a robust direct test for the difference between modified gravity and dark matter in galaxies.


\section*{Acknowledgements}

The work of MR and AH is supported by the Ferdowsi University of Mashhad. VC acknowledges the support provided by ANID through the FONDECYT grants no. 3220206 and 11250723. BF and STN acknowledge funding from the European Research Council (ERC) under the European Union’s Horizon 2020 research and innovation program (grant agreement No. 834148)".


\section*{Data Availability}

Some of the observational data set used in this paper is available online\footnote{astroweb.cwru.edu/SPARC} (see Section \ref{sec:obs} for the relevant references). The other data generated in this paper will be shared on reasonable request to TK.



\bibliographystyle{mnras}
\bibliography{Disk_stability} 








\bsp	
\label{lastpage}
\end{document}